# DOES THE MOTOR CORTEX DRAW ON A WIRE PLANE?
## A DIFFEOLOGICAL MODEL OF THE TWO-THIRDS POWER LAW

PATRICK IGLESIAS-ZEMMOUR

ABSTRACT. The two-thirds power law of human motor control ($v \propto \varkappa^{-1/3}$) is geometrically equivalent to constant equi-affine speed. In classical differential geometry, however, the equi-affine metric is not a tensor: it depends on acceleration, which does not transform covariantly under arbitrary coordinate changes. To recover tensorial behavior, one must either restrict the symmetry group to the affine group or introduce an affine connection — sacrificing full diffeomorphism covariance.

This article proposes a different geometric setting. We equip the Euclidean plane with the *wire diffeology*, the smooth structure generated by all smooth curves. In this diffeological space, the equi-affine metric becomes a true covariant 3-tensor under the *full* diffeomorphism group — no restriction of symmetries, no additional structure required.

The construction is motivated by a simple fact: the motor cortex traces curves, not two-dimensional patches. Accordingly, curves are taken as primitive, echoing the motor control literature in which movements are built from a repertoire of elementary building blocks — motor primitives. The wire plane offers a geometric formalization of this idea in which the two-thirds power law emerges as a fully covariant invariant.

## INTRODUCTION

The Two-Thirds Power Law, $v = c \cdot \varkappa^{-1/3}$, expresses a robust empirical relationship observed in human motor control: the speed $v$ of a drawing movement varies inversely with the cube root of the curvature $\varkappa$ of its trajectory [2]. Geometrically, this law is equivalent to the constancy of the equi-affine speed along the curve [4, 7]. In other words, the law reveals an affine geometric invariant hidden in the kinematics of planar hand movements.

In the standard framework of differential geometry, on the Euclidean plane $\mathbf{R}^2$ with its usual smooth structure, the equi-affine arc length is not a tensor. The reason is that acceleration involves second derivatives, which do not transform covariantly under arbitrary coordinate changes without introducing additional structure — an affine connection. To recover tensorial behavior, one must restrict the symmetry group of the plane to the affine group, effectively reducing the geometry to a preferred class of local transformations. The manifold approach achieves the best it can: the equi-affine length behaves as a tensor, but at the price of restricting the symmetries of the plane.

---

This paper proposes an alternative. Instead of restricting the symmetries, we change the smooth structure of the plane itself. We replace affine geometry with *wire geometry* by equipping the plane with the *wire diffeology* — the smooth structure generated by all smooth curves. In this diffeological space, which we call the *wire plane*, the equi-affine length becomes a true tensor without any additional assumptions. No affine connection is needed. No restriction of symmetries is required. The full diffeomorphism group is preserved [11].

The wire plane offers two conceptual gains. First, the equi-affine metric — originally born as a tensor, the cube root of a covariant 3-tensor — recovers its full tensorial character under the *entire* diffeomorphism group. Second, the geometry is built from curves, making it a natural idealization of the motor primitive repertoire proposed by Flash and colleagues [5]. In this framework, curves are primitive; points are their intersections.

A simple experiment illustrates the point. Close your eyes and draw a small oval in the air. You experience the movement as a curve. The plane emerges from the curve, not the other way around. The Two-Thirds Power Law still holds under these conditions [8]. This suggests that, at the level of motor planning, curves are primary — precisely the perspective the wire plane formalizes.

The wire plane is a geometric model. Whether it corresponds to neural implementation is a separate question, and one this paper does not address.

## 1. A PRIMER ON DIFFEOLOGY

Diffeology provides a framework for describing smooth structures on spaces that may not be manifolds. The key idea is to shift focus from charts (local coordinates) to *plots* — smooth maps from Euclidean domains into the space. A diffeological space is a set together with a collection of such plots, satisfying three simple axioms.

**Definition** (Diffeology)**.** *A* diffeology *on a set* $\mathrm{X}$ *is a set* $\mathcal{D}$ *of parametrizations* $\mathrm{P} : \mathrm{U} \to \mathrm{X}$, *where* $\mathrm{U}$ *is an open subset of* $\mathbf{R}^n$ *for some* $n$*, such that:*

1. ***Covering.*** *Every constant parametrization belongs to* $\mathcal{D}$.
2. ***Locality.*** *If* $\mathrm{P} : \mathrm{U} \to \mathrm{X}$ *is such that every point of* $\mathrm{U}$ *has a neighborhood* $\mathrm{V}$ *with* $\mathrm{P}|_{\mathrm{V}} \in \mathcal{D}$*, then* $\mathrm{P} \in \mathcal{D}$.
3. ***Smooth compatibility.*** *If* $\mathrm{P} \in \mathcal{D}$ *and* $\mathrm{F} : \mathrm{V} \to \mathrm{U}$ *is smooth (in the ordinary sense), then* $\mathrm{P} \circ \mathrm{F} \in \mathcal{D}$.

*The elements of* $\mathcal{D}$ *are called* plots*. The pair* $(\mathrm{X}, \mathcal{D})$ *is a* diffeological space.

A *smooth map* $f : \mathrm{X} \to \mathrm{Y}$ between diffeological spaces is a function such that for every plot $\mathrm{P}$ of $\mathrm{X}$, $f \circ \mathrm{P}$ is a plot of $\mathrm{Y}$. A diffeomorphism is a smooth bijection with a smooth inverse.

**(a) The standard diffeology on a manifold.** If $\mathrm{M}$ is a smooth manifold, its *standard diffeology* consists of all smooth maps $\mathrm{P} : \mathrm{U} \to \mathrm{M}$ in the usual sense. This recovers the ordinary smooth structure. In particular, for $\mathbf{R}^2$, the standard diffeology includes the identity map $(x, y) \mapsto (x, y)$, which has a 2-dimensional domain.

**(b) The wire diffeology on the plane.** Now consider the same set $\mathbf{R}^2$, but with a different diffeology.

**Definition** (Wire diffeology)**.** *A parametrization* $P : U \to \mathbf{R}^2$ *is a* plot of the wire plane $\mathscr{W}$ *if, locally around every point of* $U$, *it factors through a smooth curve. That is, for each* $u_0 \in U$ *there exists an open neighborhood* $V \subset U$, *a smooth curve* $\gamma : \mathbf{R} \to \mathbf{R}^2$, *and a smooth function* $q : V \to \mathbf{R}$ *such that* $P|_V = \gamma \circ q$.

In other words, the only smooth maps into $\mathscr{W}$ are those that, locally, depend on only one parameter. A 2-dimensional plot like the identity $(x, y) \mapsto (x, y)$ is *not* allowed, because it cannot be written as $\gamma \circ q$ for any smooth curve $\gamma$. This restriction is what allows the equi-affine metric to become a tensor, as we prove in Section 2.

**(c) Why this is not a manifold.** The wire plane $\mathscr{W}$ and the ordinary plane $\mathbf{R}^2$ have the same underlying set of points, and even the same topology [11]. Yet they are different as diffeological spaces, their smooth structures are different. A key invariant is the *diffeological dimension*: the smallest $n$ such that the diffeology can be generated by plots whose domains have dimension $n$ [9, §1.78]. For $\mathbf{R}^2$ with its standard diffeology, this dimension is 2. For the wire plane $\mathscr{W}$, because every plot factors through a curve (dimension 1), the dimension is 1.

Thus, the wire plane is a diffeological space of dimension 1 whose underlying topology is 2-dimensional — something impossible for a manifold, where dimension and topological dimension coincide. This is precisely the kind of geometry that is natural for a system — like the motor cortex — that accesses space only through curves.

**(d) Plots are probes.** The philosophical shift is worth emphasizing. In ordinary differential geometry, the smooth structure is defined by *charts*: local homeomorphisms from the space to $\mathbf{R}^n$. In diffeology, the smooth structure is defined by *probes*: smooth maps from Euclidean domains into the space. If the space is a manifold, the two notions coincide. But diffeology allows us to choose probes that match the physical or biological constraints of the problem. For the motor cortex, the natural probes are 1-dimensional curves — because the motor system produces trajectories, not 2-dimensional patches. The wire diffeology is simply the result of taking this observation seriously.[1]

## 2. The wire plane and the equi-affine tensor

Covariant tensors are genuine objects in diffeology. They are defined as maps $\alpha$ that associate to every plot $P : U \to X$ a smooth covariant tensor $\alpha(P)$ on the Euclidean domain $U$, satisfying the chain-rule condition:

$$\alpha(P \circ F) = F^*\big(\alpha(P)\big)$$

for every smooth parametrization $F$ in the domain of $P$ [9, §6.28 Note]. This definition recovers the standard notion of a tensor field when $X$ is a manifold.

[1]Diffeology has found applications beyond differential geometry. For instance, the probabilist Boris Tsirelson used it as early as 2006 to handle singular quotient structures arising from Brownian local minima, relying on an early “samizdat” version of the foundational text [9] [6].

We now define the central object of this paper: the tensor on the wire plane that makes the equi-affine length intrinsic. Since every plot of the wire plane factors locally through a smooth curve, it suffices to define the tensor on curves (the generating family) and verify that the definition respects reparametrization.

**Definition** (Equi-affine tensor on a curve)**.** *Let* $\gamma : \mathbf{R} \to \mathbf{R}^2$ *be a smooth curve. Let* Surf *denote the standard volume form on* $\mathbf{R}^2$, $\mathrm{Surf}(\mathbf{u},\mathbf{v}) = \det[\mathbf{u}\ \mathbf{v}]$. *Define a trilinear map on the tangent space of* $\mathbf{R}$ *at* $t$ *by*

$$\alpha(\gamma)_t(\delta t, \delta' t, \delta'' t) = \mathrm{Surf}\big(\dot\gamma(t)(\delta t),\ \ddot\gamma(t)(\delta' t, \delta'' t)\big),$$

*where* $\dot\gamma(t)$ *and* $\ddot\gamma(t)$ *are the first and second derivatives of* $\gamma$, *and* $\delta t, \delta' t, \delta'' t \in \mathrm{T}_t\mathbf{R} \simeq \mathbf{R}$. *For each* $\gamma$, $\alpha(\gamma)$ *is a smooth covariant 3-tensor on* $\mathbf{R}$.

The following theorem is the main mathematical result of this paper.

**Theorem.** *The family of tensors* $\alpha(\gamma)$, *defined for every smooth curve* $\gamma$ *in* $\mathbf{R}^2$, *extends uniquely to a smooth covariant 3-tensor* $\alpha$ *on the wire plane* $\mathscr{W}$.

*Proof.* First, note that $\alpha(\gamma)$ is trilinear and smooth with respect to its arguments. Thus, $\alpha(\gamma)$ is a smooth covariant 3-tensor on the domain where $\gamma$ is defined. To prove it descends to the wire diffeology, we must apply the descent criterion for differential tensors defined on generating families [9, §6.41].

We must verify that for any two local decompositions of a plot P, say $\mathrm{P} = \gamma \circ q$ and $\mathrm{P} = \gamma' \circ q'$ (as illustrated in the following figure), the pullback tensors satisfy the compatibility condition:

$$q^*(\alpha(\gamma)) = q'^*(\alpha(\gamma')).$$

Let $r \in \mathscr{U}$ be a point in the domain of the plot, and let $\delta r, \delta' r, \delta'' r \in \mathbf{R}^n$ be tangent vectors. The pullback is given by:

$$q^*(\alpha(\gamma))_r(\delta r, \delta' r, \delta'' r) = \alpha(\gamma)_t(\delta t, \delta' t, \delta'' t),$$

with $t = q(r)$ and $\delta t = \mathrm{D}(q)(r)(\delta r)$, and *mutatis mutandis* for $\delta' t$ and $\delta'' t$.

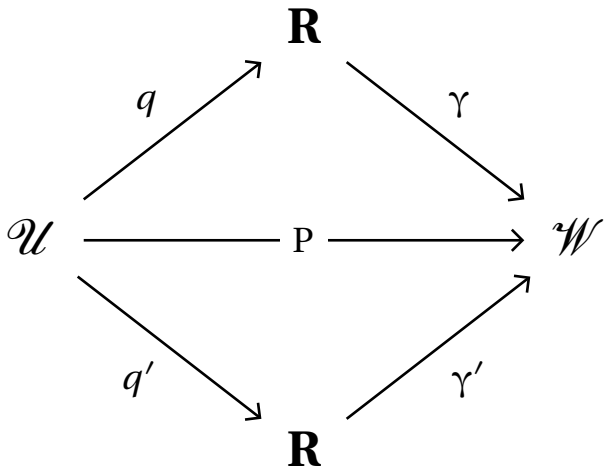

**Diagram: Independence of Parametrization.** The diagram illustrates that the measurement of the tensor does not depend on which curve ($\gamma$ or $\gamma'$) is used to represent the plot. This compatibility condition is a requirement for the geometry to be intrinsic.

Since $\gamma \circ q = \gamma' \circ q'$, differentiating once yields the collinearity of the tangent vectors:

$$\dot\gamma(t)\delta t = \dot\gamma'(t')\delta t'. \qquad (\diamondsuit)$$

Differentiating a second time, we apply the chain rule for second derivatives:

$$\mathrm{D}^2(\gamma\circ q)(r)(\delta' r)(\delta'' r) = \ddot\gamma(t)(\delta' t)(\delta'' t) + \dot\gamma(t)\,\mathrm{D}^2(q)(r)(\delta' r)(\delta'' r).$$

Equating this with the expansion for $\gamma'\circ q'$, we have:

$$\begin{aligned}\ddot\gamma(t)(\delta' t)(\delta'' t) &+ \dot\gamma(t)\,\mathrm{D}^2(q)(r)(\delta' r)(\delta'' r)\\ &= \ddot\gamma'(t')(\delta' t')(\delta'' t') + \dot\gamma'(t')\,\mathrm{D}^2(q')(r)(\delta' r)(\delta'' r).\end{aligned}$$

We now substitute these expressions into the definition of $\alpha$. We calculate the value of the tensor for the primed decomposition:

$$\begin{aligned}\mathrm{Surf}\Big(\dot\gamma'(t')\delta t',\, \ddot\gamma'(t')(\delta' t')(\delta'' t')\Big) = \mathrm{Surf}\Big(&\dot\gamma(t)\delta t,\ \ddot\gamma(t)(\delta' t)(\delta'' t)\\ &+\dot\gamma(t)\,\mathrm{D}^2(q)(r)(\delta' r)(\delta'' r)\\ &-\dot\gamma'(t')\,\mathrm{D}^2(q')(r)(\delta' r)(\delta'' r)\Big).\end{aligned}$$

By linearity of Surf in the second argument, this splits into the main term plus terms involving the second derivatives of the parametrizations. Specifically, the extra terms are of the form:

$$\mathrm{Surf}\Big(\dot\gamma(t)\delta t,\, \dot\gamma(t)\cdot \mathrm{D}^2(q)(r)(\delta' r)(\delta'' r)\Big)$$

and

$$\mathrm{Surf}\Big(\dot\gamma(t)\delta t,\, -\dot\gamma'(t')\cdot \mathrm{D}^2(q')(r)(\delta' r)(\delta'' r)\Big).$$

However, because of condition ($\diamondsuit$), the vectors $\dot\gamma(t)$ and $\dot\gamma'(t')$ are collinear. Since the volume form Surf is alternating, $\mathrm{Surf}(\mathbf{u},\mathbf{v}) = 0$ whenever $\mathbf{u}$ and $\mathbf{v}$ are collinear.

Therefore, these extra terms vanish identically. We are left with:

$$\mathrm{Surf}\Big(\dot\gamma(t)(\delta t),\, \ddot\gamma(t)(\delta' t)(\delta'' t)\Big) = \mathrm{Surf}\Big(\dot\gamma'(t')(\delta t'),\, \ddot\gamma'(t')(\delta' t')(\delta'' t')\Big).$$

This proves that $q^*(\alpha(\gamma)) = q'^*(\alpha(\gamma'))$. Consequently, $\alpha$ descends to a unique smooth covariant 3-tensor on the wire plane. □

The equi-affine arc length element $\sqrt[3]{|\alpha|}$ is therefore an intrinsic geometric quantity (a 1-density) on the wire plane, derived from the covariant 3-tensor $\alpha$. We will continue to call $\alpha$ the **equi-affine tensor**, even though no affine connection or restriction of symmetries is needed for its definition. Thus, on the wire plane, the equi-affine length transforms as a tensor under the full diffeomorphism group — a property the standard smooth structure cannot achieve without restricting symmetries or adding a connection.

**Remark.** *The construction generalizes to* $\mathbf{R}^n$ *for any* $n \geq 2$*. For a curve* $\gamma$ *in* $\mathbf{R}^n$*, define*

$$\alpha_n(\gamma)_t = \mathrm{Vol}_n\Big(\dot\gamma(t), \ddot\gamma(t), \ldots, \gamma^{(n)}(t)\Big),$$

*where* $\mathrm{Vol}_n$ *is the standard volume form on* $\mathbf{R}^n$*. The same descent argument applies: the volume form vanishes when its arguments are linearly dependent, which handles the extra terms from reparametrization exactly as in the* $n = 2$ *case. The resulting* $\alpha_n$ *is a covariant tensor of degree* $n(n+1)/2$ *on the wire space* $\mathscr{W}_n$ *(see* [1] *for the classical theory of such invariants).*

## 3. Discussion

The mathematical result of this paper is simple and, in retrospect, perhaps even obvious: if one takes seriously the fact that the motor cortex draws curves — and only curves — then the natural smooth structure on the plane is not the usual 2-dimensional manifold structure, but the wire diffeology generated by all smooth curves. In this diffeological space, which we call the wire plane, the equi-affine length underlying the Two-Thirds Power Law becomes a true covariant 3-tensor. No affine connection is needed. No restriction of symmetries is required.

This result should be compared with the standard geometric interpretation of the Two-Thirds Power Law. In the usual framework, the equi-affine length is not a tensor. To make it behave like one, one must either fix a preferred coordinate system or add an affine connection [4, 7]. Both options introduce additional structure that is not directly observable. Worse, they sacrifice full diffeomorphism covariance: the equi-affine length transforms as a tensor only under the affine subgroup, not under arbitrary smooth coordinate changes.

The wire plane restores this covariance. On the wire plane, the equi-affine length transforms as a tensor under the *full* diffeomorphism group. This is what one expects from a well-posed geometric invariant — that it should not depend on a preferred class of coordinate systems. The wire plane achieves this not by adding structure, but by choosing a smooth structure built from curves.

This choice aligns with a well-established idea in motor control research: the motor system builds complex movements from a *repertoire* of elementary building blocks — motor primitives — which can be combined and transformed according to syntactic rules [5, 10]. The wire plane provides a direct geometric formalization of this idea. If the repertoire consists of curves (strokes), then the natural smooth structure on the plane is the one generated by those curves: the wire diffeology. The Two-Thirds Power Law then emerges not as an additional constraint, but as an intrinsic tensorial property of the geometry induced by the repertoire itself. In this sense, the wire plane can be seen as an idealization of the motor primitive repertoire — a mathematical shadow cast by the curves the motor system has learned and stored.

It is worth noting that the wire plane and the ordinary plane share the same underlying set of points, the same topology, and even the same abstract group of diffeomorphisms [11]. Their difference lies entirely in their diffeology — that is, in which parametrizations are considered smooth. The wire diffeology is strictly finer: it admits fewer smooth maps. This is precisely what makes the equi-affine tensor well-defined.

The construction generalizes to higher dimensions. For spatial movements ($n = 3$), the invariant involves the third derivative (jerk). This suggests a possible connection with minimum jerk models [3], but a detailed exploration of this connection lies beyond the scope of this paper.

Finally, a word about interpretation. This paper offers a mathematical model. It does not claim that the brain “implements” diffeology or that the motor cortex “is” a wire plane. The model is mathematically rigorous, and it formalizes the observable fact that the motor cortex draws curves in a way that recovers full tensorial covariance for the equi-affine invariant. Whether biologists find this model useful — whether it helps to design new experiments or to reinterpret existing data — is an empirical question that lies outside the mathematics.

**Conclusion.** The Two-Thirds Power Law finds a natural geometric expression as an intrinsic tensor on the wire plane — one that transforms covariantly under the full diffeomorphism group, with no restriction to affine transformations. This example suggests that diffeology may offer a useful framework for modeling geometric phenomena in motor control, where the natural probes are 1-dimensional trajectories.

The present paper has considered the full wire plane, generated by all smooth curves, in order to establish the foundational result. But the motor system does not use all curves — it uses a repertoire. In diffeological terms, a motor repertoire is a generating family; the geometry it induces is the sub-diffeology generated by that family. The equi-affine tensor remains intrinsic (by pullback), concatenation of primitives is an operation within the generated diffeology, and learning corresponds to enlarging the generating family. This perspective is formulated as a set of open questions in the author's repository,[2] where it awaits further mathematical development and, perhaps, experimental collaboration.

## Acknowledgements

I thank the Hebrew University of Jerusalem for its continuous academic support. I am deeply indebted to Tamar Flash and Daniel Bennequin (private communication). It is through them that I became acquainted with this subject, and our discussions were the starting point of my reflection on the equi-affine arc length.

I am grateful to the conceptors and developers of the AI assistants Gemini (Google) and DeepSeek (深度求索), whose tools were of great help in the preparation of this article.

## References

[1] **W. Blaschke**, *Vorlesungen über Differentialgeometrie II: Affine Differentialgeometrie*, Springer, Berlin, 1923.

[2] **F. Lacquaniti, C. Terzuolo and P. Viviani**, *The law relating the kinematic and figural aspects of drawing movements*, Acta Psychologica, **54**(1-3) (1983), p. 115–130.

[3] **T. Flash and N. Hogan**, *The coordination of arm movements: an experimentally confirmed mathematical model*, Journal of Neuroscience, **5**(7) (1985), p. 1688–1703.

[4] **F.E. Pollick and G. Sapiro**, *Constant affine velocity predicts the 1/3 power law of planar motion perception and generation*, Vision Research, **37**(3) (1997), p. 347–353.

[5] **T. Flash and B. Hochner**, *Motor primitives in vertebrates and invertebrates*, Current Opinion in Neurobiology, **15**(6) (2005), p. 660–666.

[6] **B. Tsirelson**, *Brownian local minima, random dense countable sets and random equivalence classes*, Electronic Journal of Probability, **11** (2006), p. 162–298.

[7] **T. Flash and A. Handzel**, *Affine differential geometry analysis of human arm movements*, Biological Cybernetics, **96** (2007), p. 577–601.

[8] **M. Karklinsky and T. Flash**, *Timing of continuous motor imagery: the two-thirds power law originates in trajectory planning*, Journal of Neurophysiology, **113**(7) (2015), p. 2490–2499.

[2] In *GitHub Diffeology-Archives* at `https://github.com/p-i-z/Diffeology-Archives`.

[9] **P. Iglesias-Zemmour**, *Diffeology*, Mathematical Surveys and Monographs, **185**, AMS, Providence, RI, 2013.

[10] **T. Flash**, *Brain Representations of Motion Generation and Perception: Space-Time Geometries and the Arts*. In: T. Flash and A. Berthoz (eds), *Space-Time Geometries for Motion and Perception in the Brain and the Arts*, Lecture Notes in Morphogenesis, Springer, Cham, 2021, p. 3–34.

[11] **P. Iglesias-Zemmour**, *The Boman paradox: When topology plus algebra do not define geometry*, The Mathematical Intelligencer, 2026. https://doi.org/10.1007/s00283-025-10500-3

**Note on Data.** Data sharing is not applicable to this article as no new data were created or analyzed in this study.

EINSTEIN INSTITUTE OF MATHEMATICS, THE HEBREW UNIVERSITY OF JERUSALEM, CAMPUS GIVAT RAM, 9190401 ISRAEL

*Email address*: piz@math.huji.ac.il